\documentclass[10pt,prd,longbibliography,superscriptaddress]{revtex4-2}

\usepackage[T1]{fontenc}
\usepackage[utf8]{inputenc}
\usepackage[english]{babel}
\usepackage{gensymb}

\usepackage{amsfonts,amsbsy,amssymb,amsmath,graphicx,float} 
\usepackage{color}

\usepackage{textcomp}
\usepackage{gensymb}

\graphicspath{{figures/}}

\usepackage[linktoc=all,colorlinks=true,citecolor=blue,linkcolor=blue]{hyperref}

\begin{document}


\title{The time evolution of the trajectories after the selectivity in  a symmetric potential energy surface with a post-transition-state bifurcation}

\author{\firstname{Douglas}~\surname{Haigh}}%
\email[E-mail: ]{jl18114@bristol.ac.uk}
\affiliation{
School of Mathematics, University of Bristol, \\ Fry Building, Woodland Road, Bristol, BS8 1UG, United Kingdom.}%

\author{\firstname{Matthaios}~\surname{Katsanikas}}%
\email[E-mail: ]{matthaios.katsanikas@bristol.ac.uk}
\affiliation{
School of Mathematics, University of Bristol, \\ Fry Building, Woodland Road, Bristol, BS8 1UG, United Kingdom.}%

\author{\firstname{Makrina}~\surname{Agaoglou}}%
\email[E-mail: ]{makrina.agaoglou@bristol.ac.uk}
\affiliation{
School of Mathematics, University of Bristol, \\ Fry Building, Woodland Road, Bristol, BS8 1UG, United Kingdom.}%

\author{\firstname{Stephen}~\surname{Wiggins}}%
\email[E-mail: ]{s.wiggins@bristol.ac.uk}
\affiliation{
School of Mathematics, University of Bristol, \\ Fry Building, Woodland Road, Bristol, BS8 1UG, United Kingdom.}%
\begin{abstract}
   Selectivity is an important phenomenon in chemical reaction dynamics. This can be quantified by the branching ratio of the trajectories that visit one or the other wells to the total number of trajectories in a system with a potential with two sequential  index-1 saddles and two wells (top well and bottom well). In our case, the relative branching ratio is 1:1 because of the symmetry of our potential energy surface. The mechanisms of transport and the behavior of the trajectories in this kind of systems have been studied recently. In this paper we study the time evolution after the selectivity as energy varies using periodic orbit dividing surfaces. We investigate what happens after the first visit of a trajectory to the region of the top or the bottom well for different values of energy. We answer the natural question, what is the destiny of these trajectories?
\end{abstract}

\maketitle


\section{Introduction}
\label{intro}

\begin{figure}[htbp]
	\begin{center}
        \includegraphics[scale=0.30]{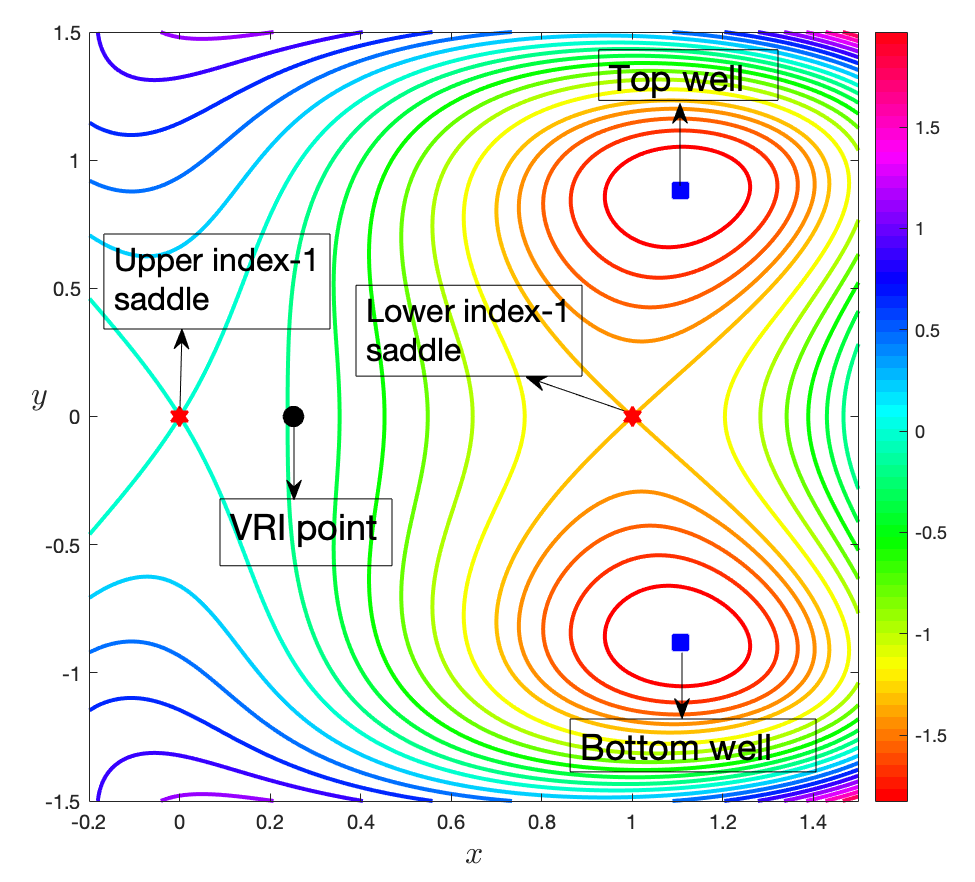} 
	\end{center}
	\caption{Equipotential contour plot of the PES. The red asterisks represent the index-1 saddle points. The blue squares represent the two potential wells (top well and bottom well).}
	\label{VRI}
\end{figure}

In this paper we are concerned with the global dynamical properties of a two degree-of-freedom (DoF) Hamiltonian system having a potential energy surface (PES) with four critical points: two index-1 saddles and two minima. One index-1 saddle (the upper index-1 saddle) has higher energy than the other index-1 saddle (the lower index-1 saddle) and the lower index-1 saddle separates the two potential wells (the upper well and the lower well). Between the two index-1 saddles is a valley ridge inflection (VRI) point (where the shape of the potential near the VRI point changes from a valley to a ridge). We will provide an analytical description of the PES in Section \ref{model} , but in Fig. \ref{VRI} we plot the equipotential contours of the PES that we will study showing the critical points as described above. The PES that we study in this paper is symmetric in the sense that we describe precisely in Section \ref{model}.

\noindent

Motivation for studying the dynamics associated with this form of PES come from the topic of organic chemical reactions \cite{Valtazanos1986, singleton2003, carpenter2004potential, carpenter2005nonstatistical, ess2008bifurcations, thomas2008control, birney2010, rehbein2011we, hare2017post}. The index-1 saddle points are considered to be ``transition states'', i.e. points on the PES that denote that ``transition'' from reactants to products. The wells denote the products of the reaction. As a trajectory crosses the high energy index-1 saddle in the direction of the low energy index-1 saddle it exhibits three possible behaviors (at least, on short to medium time scales). It can enter the upper well, the lower well, or it can avoid both wells and reflect off the back of the right-hand portion of the PES. In the chemistry literature this ``choice'' that trajectories have after crossing the high energy index-1 saddle in this way is referred to as a {\em post transition state bifurcation} (see \cite{collins2013nonstatistical}). 

Of interest is which wells are first visited by trajectories that cross the high energy index-1 saddle. An ensemble of trajectories of fixed total energy  that cross the high energy index-1 saddle are initiated at the saddle and the number that visit either the top or bottom well is counted.   The ratio of the  number of trajectories that  visit the top well to those that visit the bottom well is known as the branching ratio. The characterization of trajectories in terms of which well they first visit is referred to as {\em selectivity}. 

Detailed trajectories studies of the branching ratio and product selectivity for a two DoF PES of the type described here were carried out in \cite{collins2013nonstatistical}. That particular PES was more complicated than the one that we consider here in that it had more parameters that allowed for the consideration of both symmetric and asymmetric cases. Our focus on the symmetric case allows us to focus on the phase space mechanisms that govern product selectivity since in the symmetric case we know that 
$50\%$ of the trajectories first enter the top well and $50\%$ of the trajectories first enter the  bottom well. Hence, the symmetry of the potential is reflected in a 1:1 branching ratio.
The phase space mechanisms responsible for this in the symmetric case were studied in 
\cite{Agaoglou2020,katsanikas2020c}. We remark that the 1:1 branching ratio can be altered to any chosen value by the choice of an appropriate external time dependence. This was shown in \cite{garcia2020tuning}.

It is natural to consider the time evolution of trajectories beyond their first visit to a well, i.e., the time evolution after selectivity. Likewise, the question of the nature of the dependence of these dynamical processes on the total energy of the system occurs simultaneously. These issues are the topic of this paper.

The structure of the paper is as follows. In section \ref{model} we describe the PES and we present the location and the total energy of the equilibrium points and the VRI point. 
In section \ref{sec.1} we present the families of periodic orbits from which periodic orbit dividing surfaces (PODS) are constructed in order to quantify selectivity and its time evolution. Section \ref{sec.2} describes our results and conclusions are given in Section \ref{concl.}.

\section{Model}
\label{model}

In Section \ref{intro} we described the PES whose associated dynamics we will analyze in relation to the concept of product selectivity. In this section we will consider the analytical form of the PES and the associated Hamilton's equation. 

The PES that we use was inspired by that given in \cite{collins2013nonstatistical}:

\begin{equation}
V(x,y) = \frac{8x^3}{3} - 4x^2 + \frac{y^2}{2} + xy^2(y^2-2).
\label{PES}
\end{equation}

\begin{figure}[htbp]
	\begin{center}
		\includegraphics[scale=0.3]{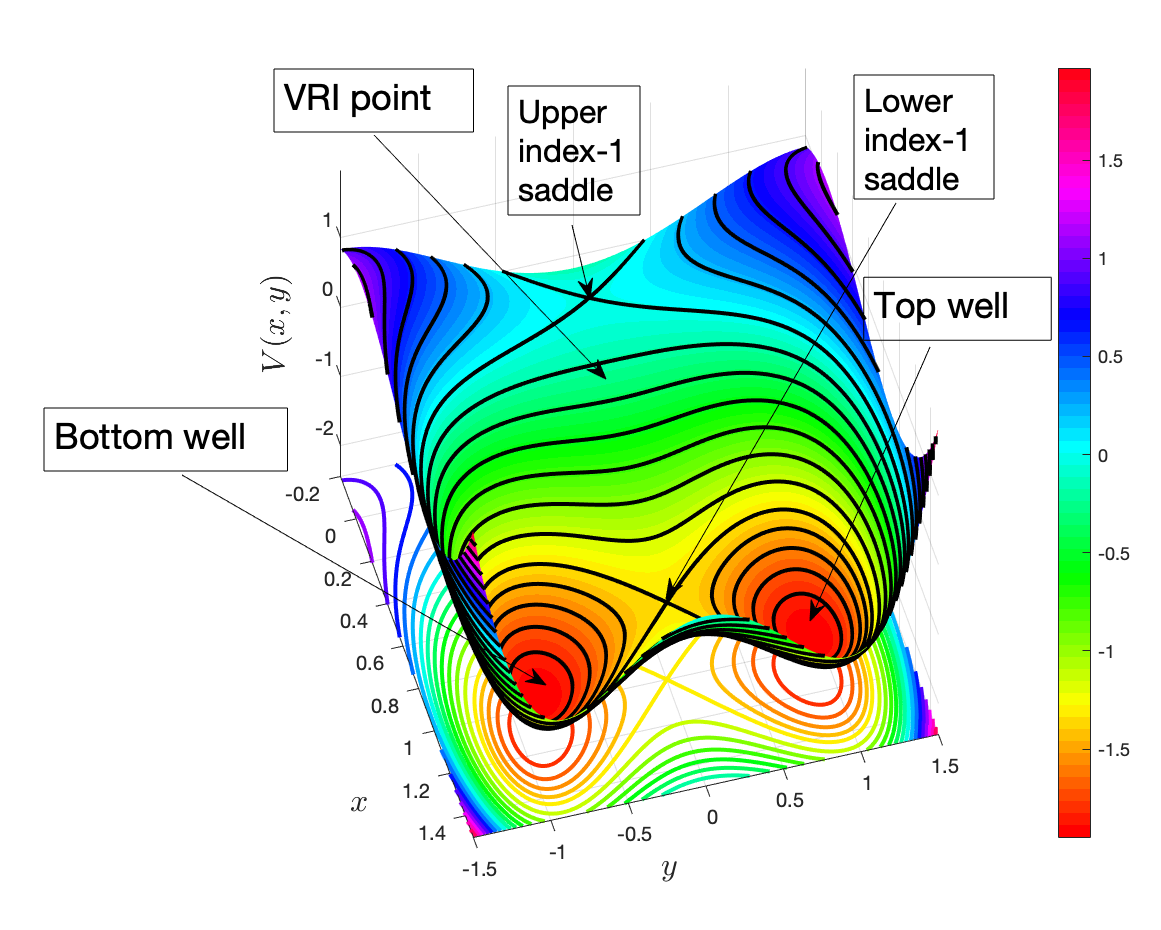} 
	\caption{Plot of the PES given in Eq. (\ref{PES}).}
    \end{center}
    \label{pes}
\end{figure}

\noindent
As we noted in Section \ref{intro} and illustrated in Figure \ref{VRI}, the PES has an upper index-1 saddle (exit/entrance channel) and a lower index-1 saddle, which is an energy barrier separating two potential wells. Moreover, the PES is symmetric about the x axis. A three dimensional plot of the PES is given in Figure \ref{pes}.

The location of the critical points of the PES Eq. (\ref{PES}), together with their linear stability and their energies can be found in the Table \ref{tab:tab1}. We note that the VRI point lies on the $x$-axis at the location $(1/4,0)$ and has total energy is $-5/24$.

\begin{table}[htbp]	
	\begin{tabular}{| l | c | c | c|}
		\hline
		Critical point \hspace{1cm} & \hspace{0.6cm} Location $(x,y)$ \hspace{0.6cm} & \text{Energy} $(V)$ & \hspace{.6cm} \text{Stability} \hspace{.6cm} \\
		\hline\hline
		Index-1 Saddle (Upper) & $(0,0)$ & 0 & saddle $\times$ center \\
		\hline
		Index-1 Saddle (Lower) & $(1,0)$ & -4/3 & saddle $\times$ center \\
		\hline
		Potential Well (Top) & $(1.107146,0.879883)$ & -1.94773 & center \\
		\hline
		Potential Well (Bottom) & $(1.107146,-0.879883)$ & -1.94773 & center  \\
		\hline
		\end{tabular} 
		\caption{Location and energies of the critical points of the PES, together with their linear stability behavior when considered as equilibrium points of Hamilton's equations.} 
	\label{tab:tab1} 
\end{table}

\noindent
Finally, our interest is in the dynamics generated by Hamilton's equations. In regard to that, the Hamiltonian is given by

\begin{equation}\label{hamil}
  H(x,y,p_x,p_y) = \frac{p_x^2}{2} + \frac{p_y^2}{2} + V(x,y),   
\end{equation}

\noindent
and the corresponding Hamiltonian vector field (i.e. Hamilton's equations of motion) are given by:

\begin{equation}\label{ham1}
 \begin{cases}
   \dot{x} = p_x,\\
   \dot{y} = p_y,\\
   \dot{p}_x = 8x(1-x) + y^2(2-y^2), \\
   \dot{p}_y = y \left(4x(1-y^2)-1 \right).
  \end{cases}
\end{equation}

\noindent
Hamilton's equations conserve the total energy, which we will denote as $E$ throughout the text.


\section{Periodic orbits and Dividing surfaces}
\label{sec.1}

In this section, we describe the families of unstable periodic orbits whose associated periodic orbit dividing surfaces control access to the top and bottom wells. Our system has two index-1 saddles, one with a higher (upper saddle) energy that the other (lower saddle). As the energy is increased above that of the lower saddle an unstable periodic orbit bifurcates from the saddle, which follows from the Lyapunov subcenter theorem (\cite{rabinowitz1982periodic,weinstein1973normal,moser1976periodic}). This means that the Lyapunov family of periodic orbits  emanating from the saddle, initially unstable, turns to stable by the pitchfork bifurcation, and two new unstable families of periodic orbits appear. 

\begin{figure}[htbp]
	\begin{center}
		A)\includegraphics[scale = 0.45]{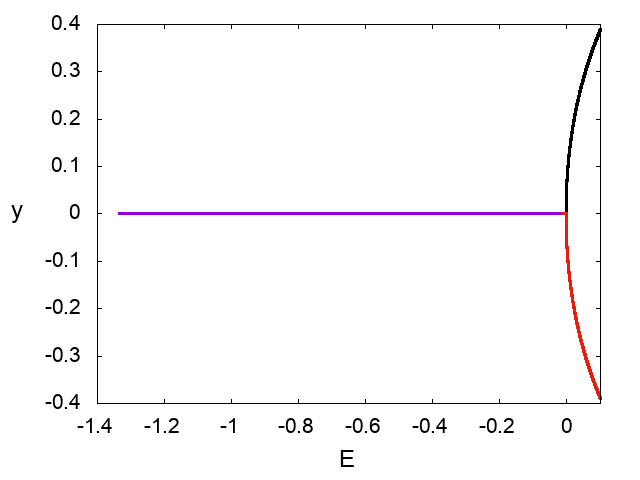} 
		B)\includegraphics[scale = 0.45]{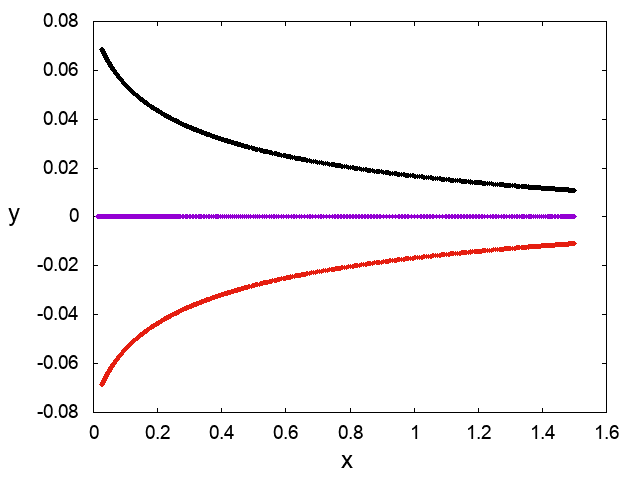}
	\end{center}
	\caption{A) The characteristic diagram (on the section $x=0.05$) of the families of periodic orbits of the lower index-1 saddle (with violet color) and their bifurcations, the families of the top (with black color) and bottom (with red color) unstable periodic orbits. B) The periodic orbits of the families of the lower index-1 saddles (with violet color) and of their bifurcations, top (with black color) and bottom (with red color) unstable periodic orbits, in the configuration space for energy $E=-0.00055$ (just after the bifurcation point).}
	\label{sdiagram}
\end{figure}

We computed this Lyapunov family of periodic orbits and followed it for increasing energy using a standard continuation method (the Newton-Raphson algorithm)  on the  section  $x=0.05$ (the  general form of the initial condition used was $(x,y,p_x,p_y)$). The evolution of this periodic orbit is depicted in the {\em characteristic diagram} which shows the evolution of one coordinate of the periodic orbit (in our case the $x$ coordinate) as a function of the energy. This is shown in panel A of Fig. \ref{sdiagram}. This family undergoes a pitchfork bifurcation at $E=-0.00056$, which is just below the energy of the upper saddle, which has energy $E=0$. This is shown in panel B of Fig. \ref{sdiagram}. 

\begin{figure}[htbp]
    \begin{center}
        A)\includegraphics[scale = 0.20]{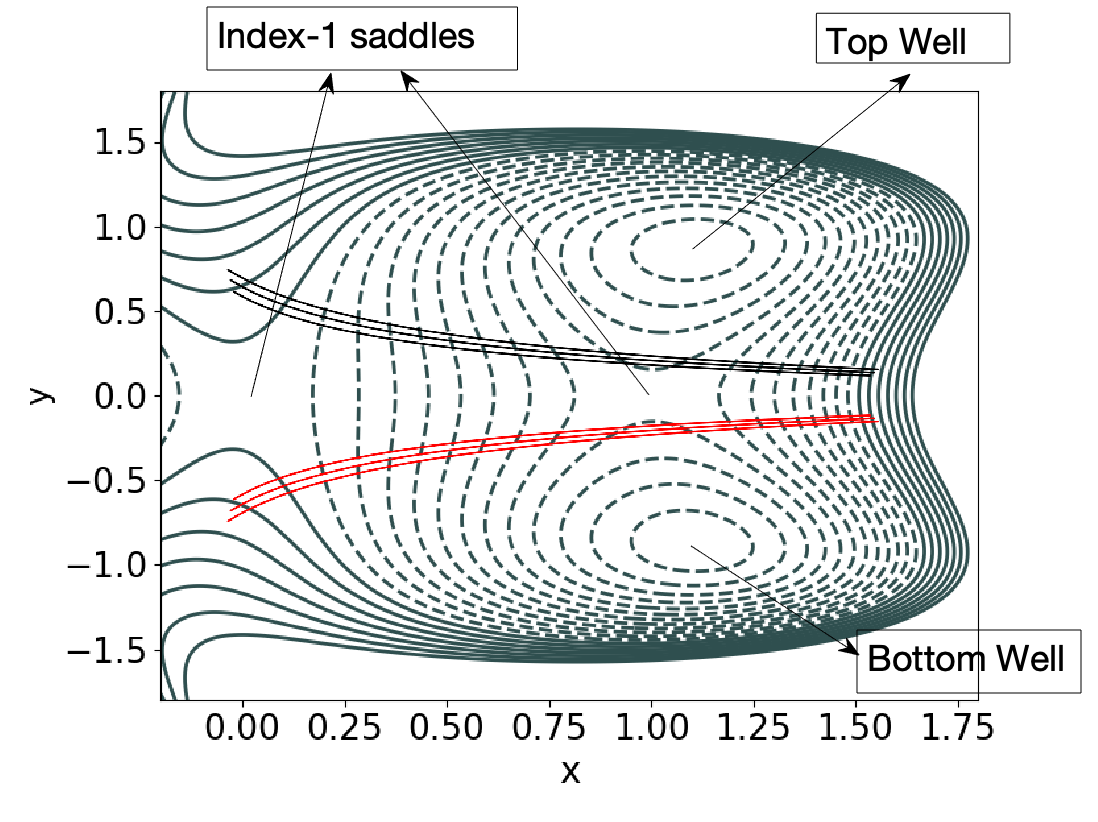}
        B)\includegraphics[scale = 0.20]{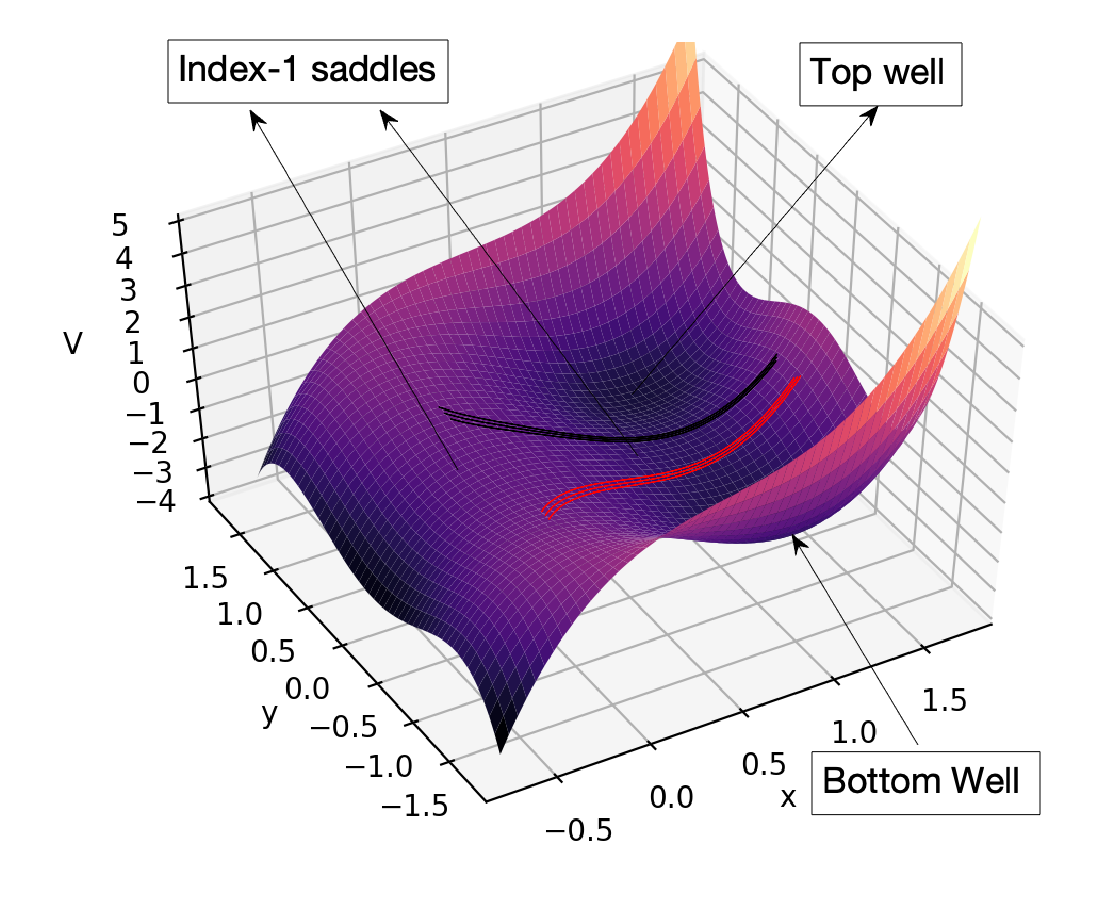}
    \end{center}
    \caption{In this figure we depict  3 top (with black color) and the 3 corresponding bottom (with red color) periodic orbits A) in the configuration space with the contours of the potential.  B) in the 3D space $(x,y,V)$.}
    \label{top1}
\end{figure}

We refer to the two new families of unstable periodic orbits created in the pitchfork bifurcation as the top and the bottom unstable periodic orbits. When we projected into the configuration space these periodic orbits lie at the borders of the region of the top and the bottom wells, as we can see in the panels A and B of Fig. \ref{top1}.

\begin{figure}[htbp]
	\begin{center}
		\includegraphics[scale = 0.35]{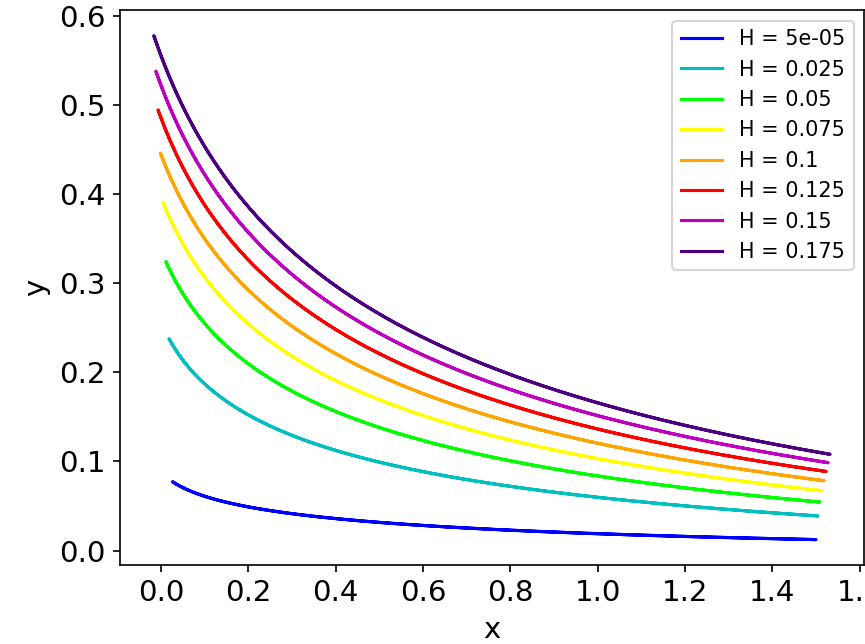} 
		\includegraphics[scale = 0.35]{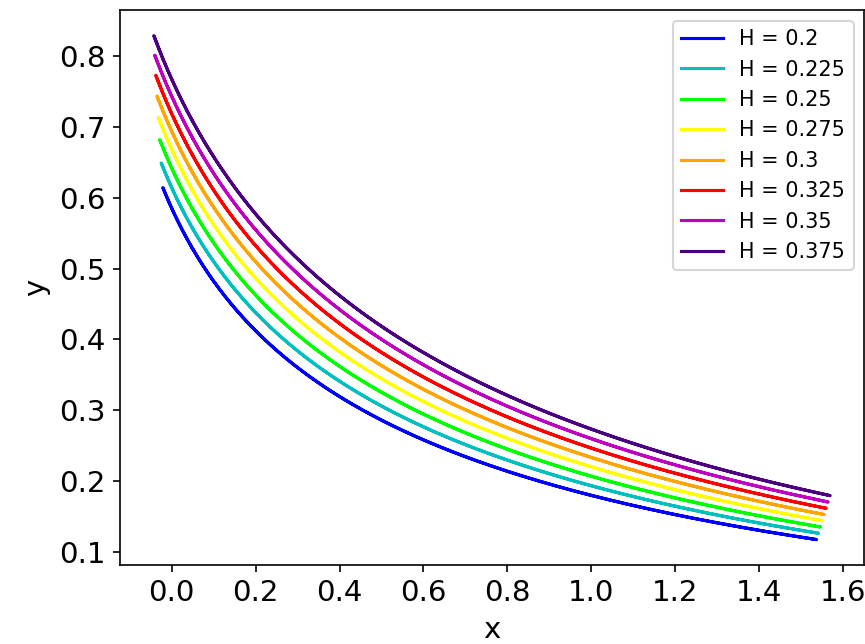}
		\includegraphics[scale = 0.35]{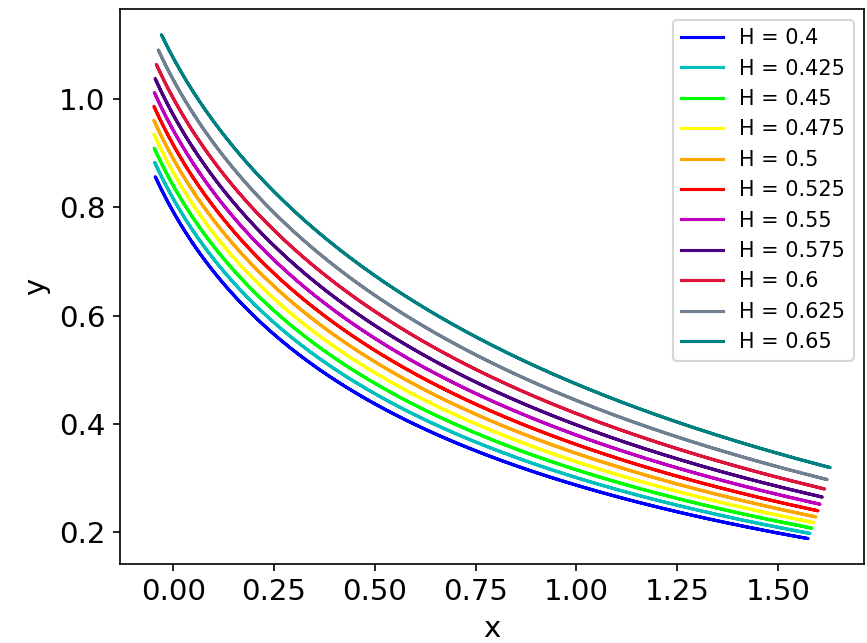}
	\end{center}
	\caption{ The periodic orbits in the $ x - y $ plane, or configuration space, with the numerical value of the Hamiltonian "H" (energy $E$) labelled for each orbit. }
	\label{poxy}
\end{figure}

\begin{figure}[htbp]
	\begin{center}
		\includegraphics[scale = 0.4]{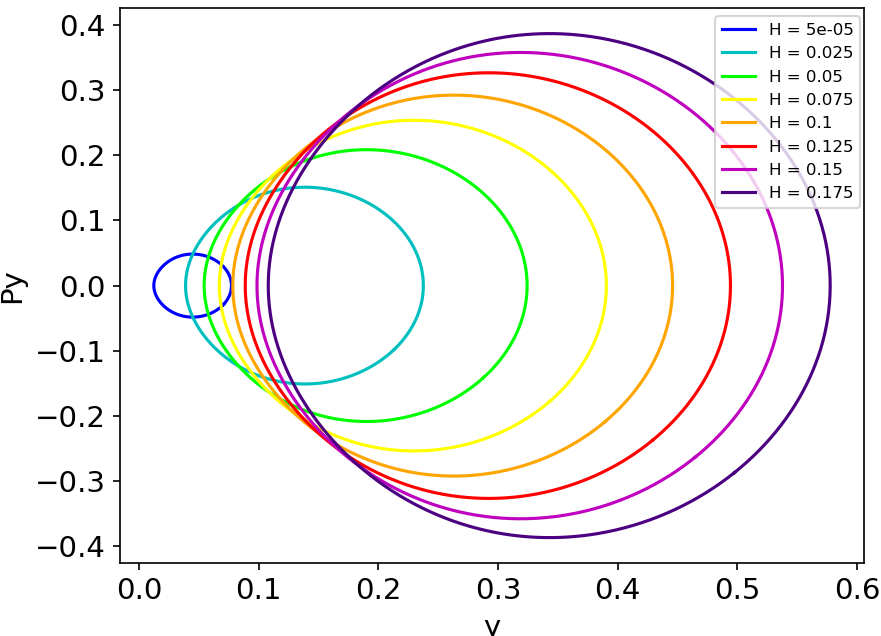} \includegraphics[scale = 0.4]{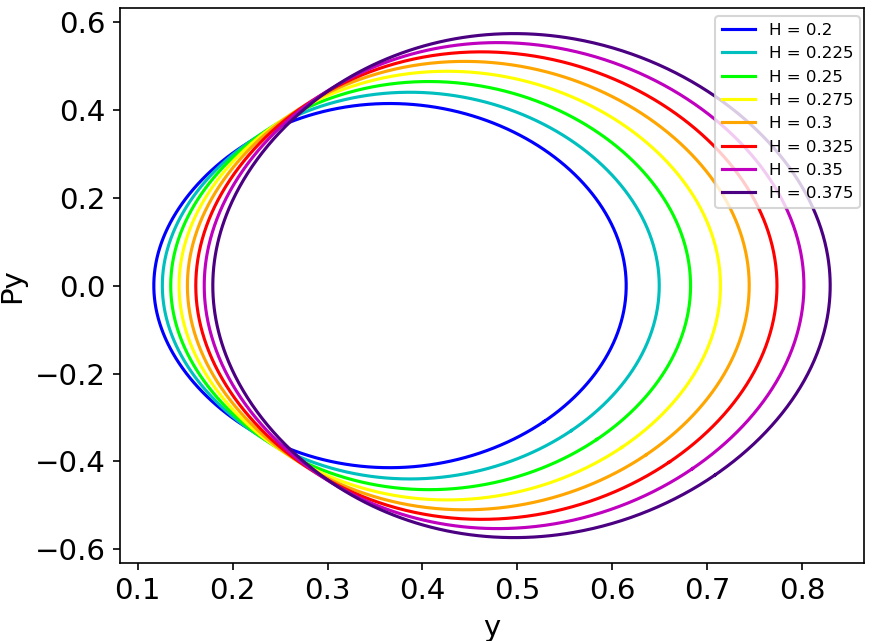}
		\includegraphics[scale = 0.4]{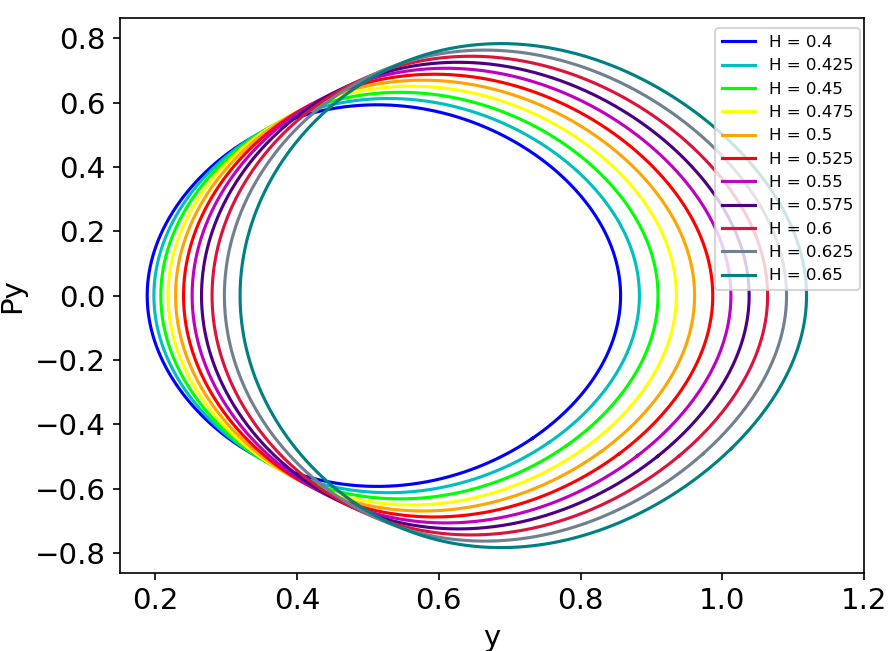}
	\end{center}
	\caption{ The periodic orbits in the $ y - p_y $ plane, with the numerical value of the Hamiltonian "H" (energy $E$) labelled for each orbit. }
	\label{poypy}
\end{figure}

\begin{figure}[htbp]
	\begin{center}
		\includegraphics[scale = 0.4]{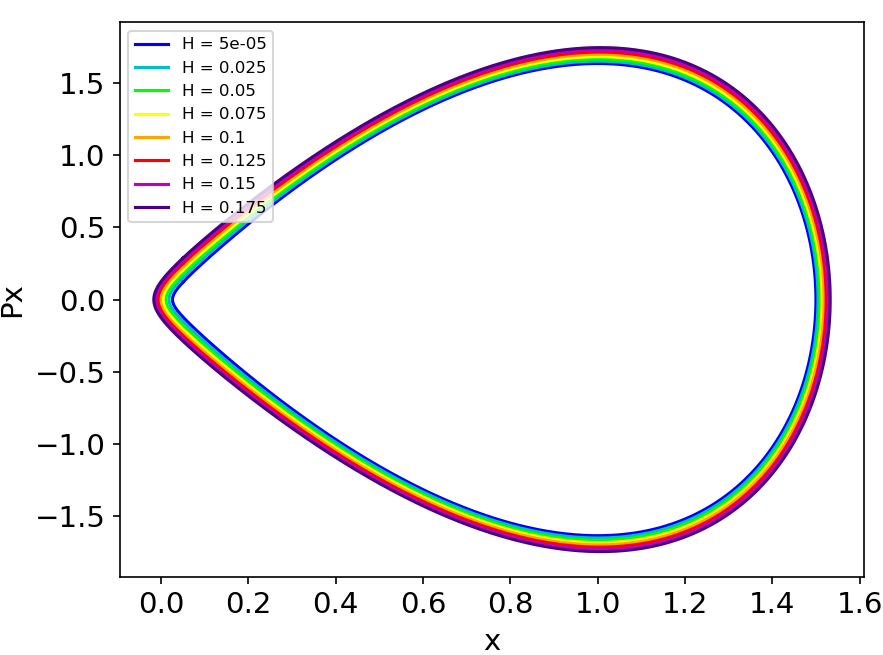} \includegraphics[scale = 0.4]{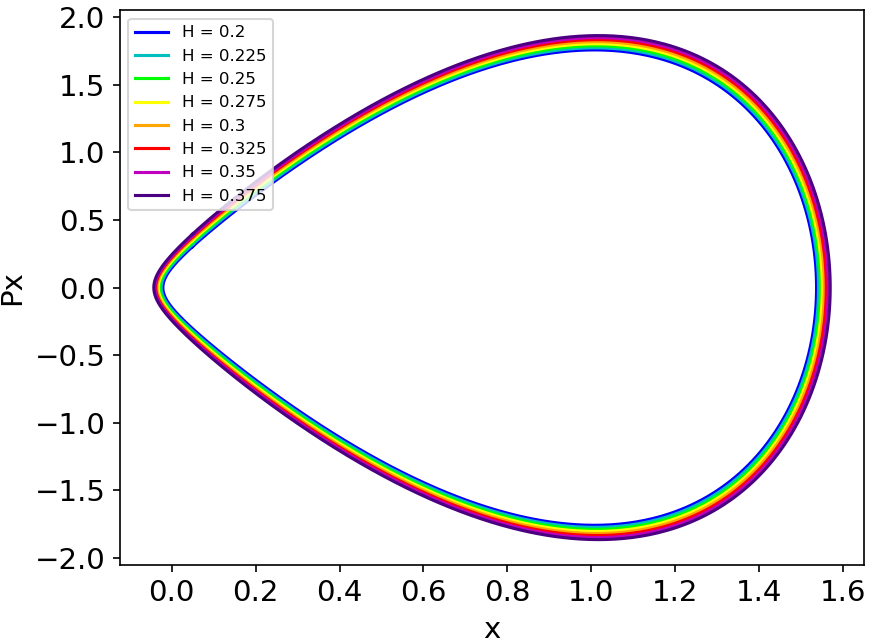}
		\includegraphics[scale = 0.4]{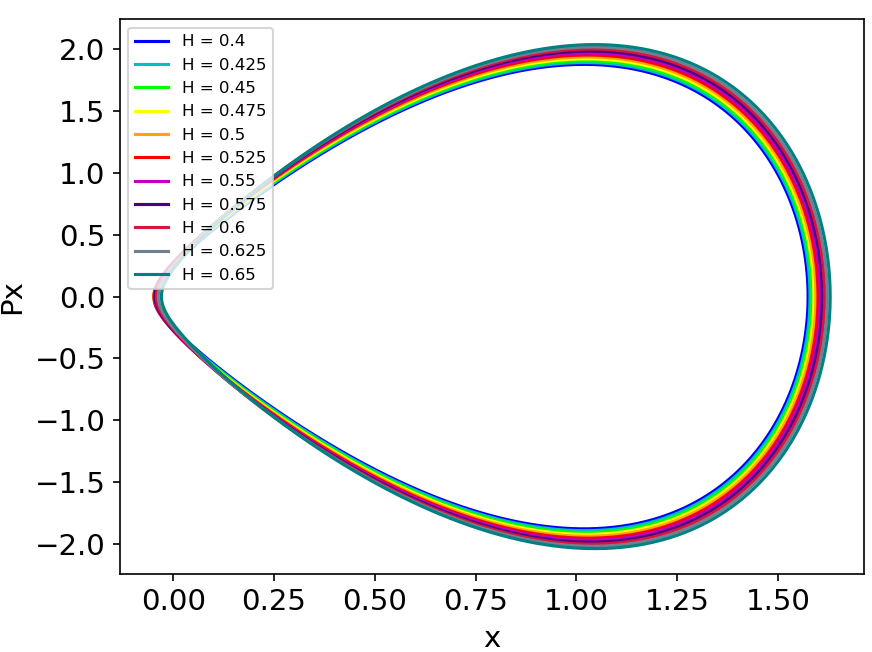}
	\end{center}
	\caption{The  periodic orbits in the $ x - p_x $ plane with the numerical value of the Hamiltonian "H" (energy $E$) labelled for each orbit. }
	\label{poxpx}
\end{figure}

We notice that as we increase the energy, the periodic orbits are more elongated in the y-axis in the configuration space (see Fig. \ref{poxy})  and they increase their range in $p_y$ (see Fig. \ref{poypy}), but they exhibit little change in the $x$-direction or $p_x$-direction (see Fig. \ref{poxpx}). Then we applied all the steps of the algorithm for the top unstable periodic orbits that are shown  in Figs. \ref{poxy}, \ref{poypy} and \ref{poxpx}.

As described in the introduction, if a trajectory in the configuration space enters into the region of the top or the bottom well its projection into configuration space must cross the projection of these periodic orbits. This means that if we construct a dividing surface in the phase space in the region of the top and bottom unstable periodic orbits we can monitor the entrance and the exit of the trajectories at the regions of the top and the bottom wells.

The methodology and utility of the construction of phase space dividing surfaces from periodic orbits for two degree-of-freedom Hamiltonian systems was described in \cite{Pechukas73,Pechukas77,Pollak78,Pechukas79}. We will use the algorithm for construction as described in \cite{waalkens2004direct, ezra2018sampling}, which we rewrite below.

\begin{enumerate} 

\item Locate a periodic orbit.

\item Project the periodic orbit into configuration space.

\item Choose points on that curve $(x_i,y_i)$ for $i=1,...N$ where  $N$ is the desired number of points.  Points are spaced uniformly according to distance along the periodic orbit. 

\item For each point  $(x_i,y_i)$ determine $p_{x max,i}$ by solving:

\begin{eqnarray}
\label{eq3a}
 H(x_i, y_i, p_x, 0)=\frac {p_x^2}{2m}+ V(x_i, y_i)
\end{eqnarray}

\noindent
for $p_x$. Note that a solution of this equation requires $E-V(x_i,y_i) \geq 0$ and that there will be two solutions, $\pm p_{x max,i}$. 
\item For each point $(x_i,y_i)$ choose points for $j=1,...,K$ with $p_{x_1}=-p_{x max,i}$ and  $p_{x_K}=p_{x max,i}$  and solve the equation  $H(x_i, y_i, p_x, p_y)=E$ to obtain $p_y$ (we will obtain two solutions  $p_y$, one negative and one positive) .

\end{enumerate}

The top and bottom unstable periodic orbits are symmetric. Consequently, results obtained for transport in and out of one well will apply to the other. Hence, we will focus on the top well and compute the periodic orbit dividing surface for the top unstable periodic orbit.

\section{The time evolution of the trajectories after selectivity}
\label{sec.2}

In this section we will investigate the time evolution of the trajectories after  selectivity for our system. As we described before (see  Section \ref{intro})  selectivity refers to the characterization of the trajectories in terms of (that start from the region of the high energy saddle) which well they first  visit. In this section, we want to examine the time evolution of these trajectories after selectivity (after the first visit to the region of  one of the wells). This means that the starting point of our investigation is just after  selectivity. It is the moment that the trajectories  enter into or exit from the region of one of the two wells. Our goal is to see the time evolution of an ensemble of trajectories (10000) after the selectivity. Recall that by symmetry half of them enter into  the region of the one well and the other half exit from this. Then we compute the ratio of these trajectories to the total number of trajectories (after a time interval of numerical integration) that visit the different regions of the configuration space: the region of the top or bottom well, the entrance/exit channel, or the exit: 

\begin{enumerate}

\item {\bf Ratio of the trajectories that are in the top well}: The number of the trajectories that are in the region of the top well after a time interval of integration over the whole set of the trajectories that we considered after  selectivity.

\item {\bf Ratio of the trajectories that are in the bottom well}: The number of the trajectories that are in the region of the bottom well after a time interval of integration over the whole set of the trajectories that we considered after  selectivity.

\item {\bf Ratio of the trajectories that are in the entrance/exit channel}: The number of the trajectories that are in the region of the entrance/exit channel after a time interval of integration over the whole set of the trajectories that we considered after the selectivity.

\item {\bf Ratio of the trajectories that exit}: The number of the trajectories that go to the exit after a time interval of integration over the whole set of the trajectories that we considered after  selectivity.

\end{enumerate}

In this section, we focus on the trajectories after selectivity 
for only the case of the top well. In a similar way we can obtain the 
results for the bottom well due to the symmetry of the potential energy surface. We see in the panel A of Fig. \ref{top1} that  the top  unstable periodic orbits (black color) are at the borders of the regions of the top  well in the configuration space. This means that if we construct the dividing surfaces (see Section \ref{sec.1}) from these periodic orbits we can divide the corresponding  regions of the top well from the other regions. 

The following steps describe our numerical simulation of trajectories after selectivity for the case of the top well:
 
\begin{enumerate}

    \item The first step is  to  choose 10000 initial conditions on the dividing surfaces of the top  unstable periodic orbits. These initial conditions have positive and negative values for the $p_y$ coordinate (see step 5 of the algorithm for the construction of the dividing surface in Section \ref{sec.1}). Half of them have  positive values of $p_y$ and  correspond to trajectories that  enter  the region of the top well.  The other half of  the initial conditions have negative $p_y$ values and  correspond to trajectories that exit from the region of the top well.

    \item The next step is to  integrate  the initial conditions on the dividing surfaces of the top  unstable periodic orbits and study the time evolution of the trajectories for a time  interval $[0,6]$. We chose this time interval  because we know  (\cite{katsanikas2020c},\cite{Agaoglou2020},\cite{garcia2020tuning}) that we need 3 time units for the transport of the trajectories from the region of the upper index-1 saddle to the regions of the wells. We wanted to have a larger time interval (we doubled this interval of time)  than this in order to have enough time for the trajectories that enter or leave the region of the wells to  be guided to the exit even if we have long times of trapping in these regions. 
    
    \item The next step is to  detect every 1/10 (0.6 time units) of the time interval of the integration the position of the trajectories. We count the trajectories that are in the region of the top well, the trajectories that are in the region of the bottom well, the trajectories that are in the region of the entrance/exit channel and the trajectories that exit. 
    According to \cite{garcia2020tuning} a trajectory visits the region of the top or the bottom well if its $y$ coordinate is larger than 0.5 (for the region of the top well) or smaller than -0.5 (for the region of the bottom well). In addition, we consider that a trajectory exits if its $x$ coordinate  is lower than 0. If a trajectory does not satisfy any of the previous conditions, this means that the trajectory is at the region of the entrance/exit channel.
    
    \item We repeat the previous steps  for many values of energy in order to see the evolution of the trajectories for the same time interval as the energy of the system increases. For our computations we used 27 values of energy in the range from 0 to 0.65 (with a step 0.25). 

\end{enumerate}

\begin{figure}[htbp]
    \begin{center}
        A)\includegraphics[scale = 0.5]{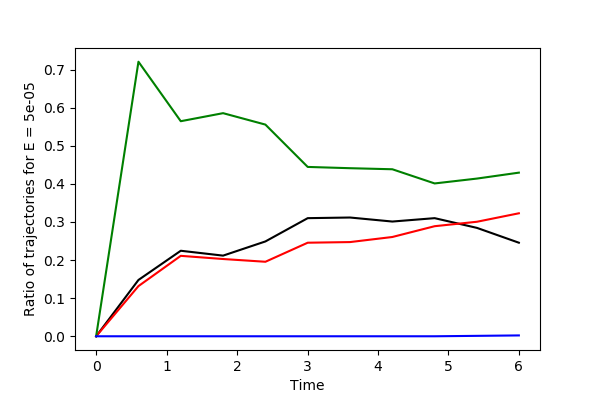}  
        B)\includegraphics[scale = 0.5]{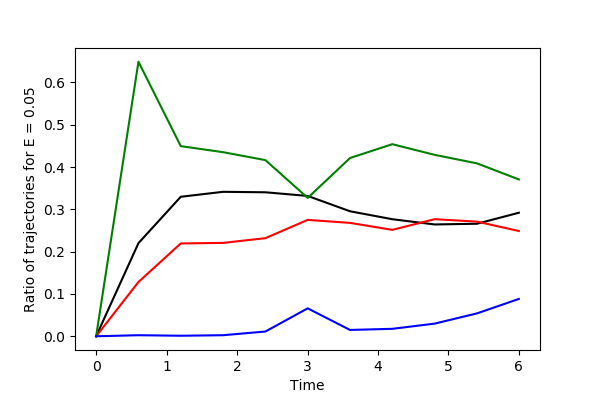}\\
        C)\includegraphics[scale = 0.5]{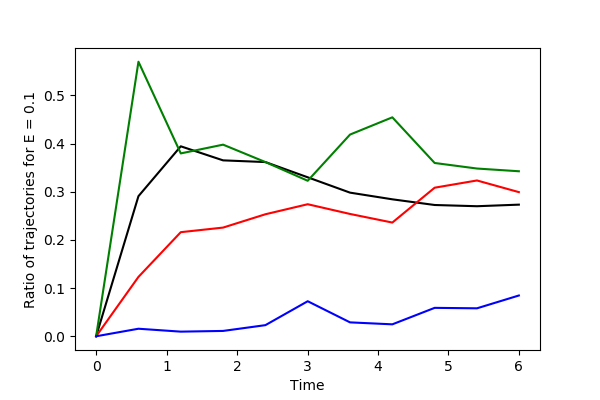}
        D)\includegraphics[scale = 0.5]{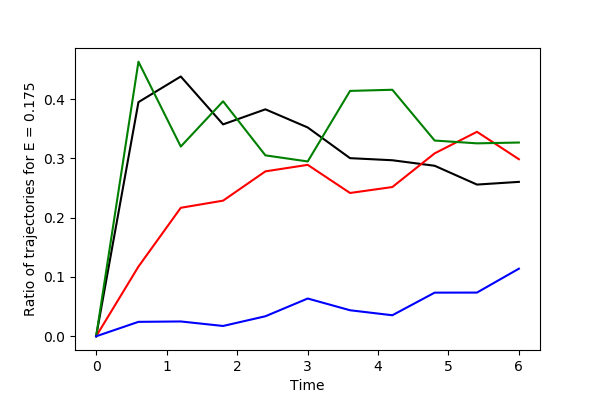}\\
     \end{center}
     \caption{ The diagrams depicting the  ratio of the trajectories that visit certain regions versus time for low values of energy. Each plot corresponds to a different value of energy. The panels A,B,C and D correspond at values of energy $5\times 10^{-5}$, $0.05$, $0.1$ and $0.175$, respectively. The black, red and  green line  in each diagram  represent the ratio of the trajectories  that visit the top well, the bottom well and  the entrance/exit channel. The blue lines represent the ratio of trajectories that exit.}
     \label{tratio1}
\end{figure}

\begin{figure}[htbp]
    \begin{center}
        A)\includegraphics[scale = 0.5]{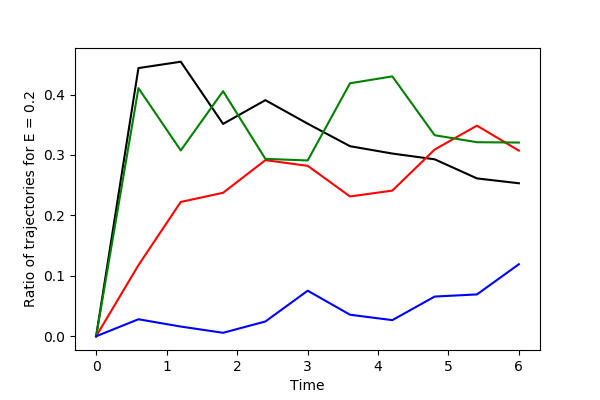}
        B)\includegraphics[scale = 0.5]{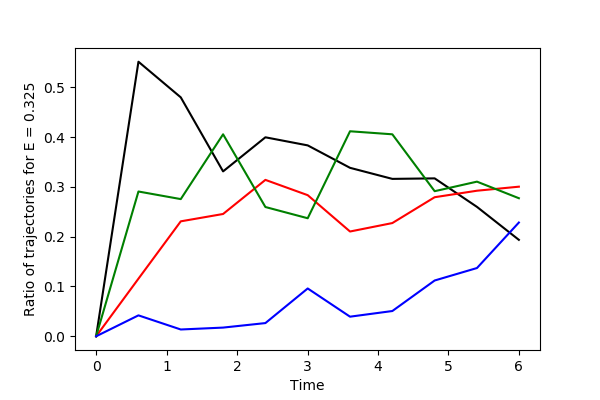}\\
        C)\includegraphics[scale = 0.5]{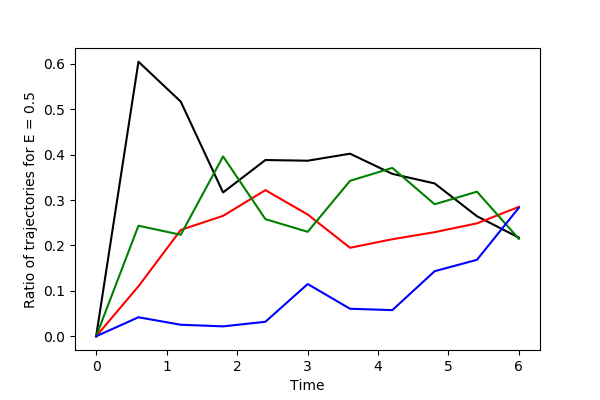}  
        D)\includegraphics[scale = 0.5]{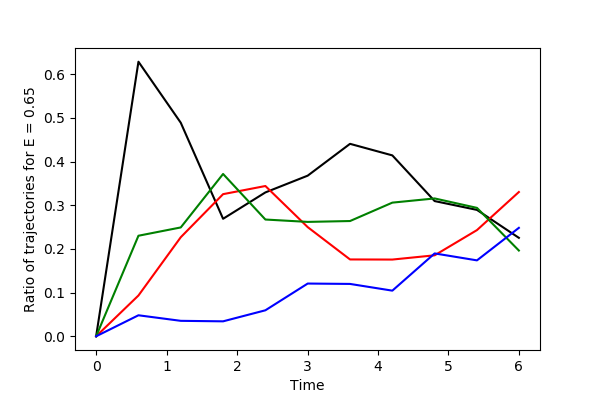}\\
        \end{center}   
    \caption{ The diagrams depicting the  ratio of the trajectories that visit certain regions versus time for high values of energy. Each plot corresponds to a different value of energy. The panels A,B,C and D correspond at values of energy $0.2$, $0.325$, $0.5$ and $0.65$, respectively. The black, red, and  green line  in each diagram  represent the ratio of the trajectories  that visit the top well, the bottom well and  the entrance/exit channel, respectively. The blue lines represent the ratio of trajectories that exit.} 
    \label{tratio2}
\end{figure}

We present the diagrams that show the ratio of the trajectories that visit the top well (black line), the bottom well (red line), the entrance/exit channel (green line) and the trajectories that exit (blue line) in a  time interval $[0,6]$ for different values of energy in Fig. \ref{tratio1} and \ref{tratio2}. In particular  Fig. \ref{tratio1} is for energies below 0.2 and  Fig. \ref{tratio2} is for energies above 0.2.

\begin{figure}[htbp]
    \begin{center}
        A)\includegraphics[scale = 0.5]{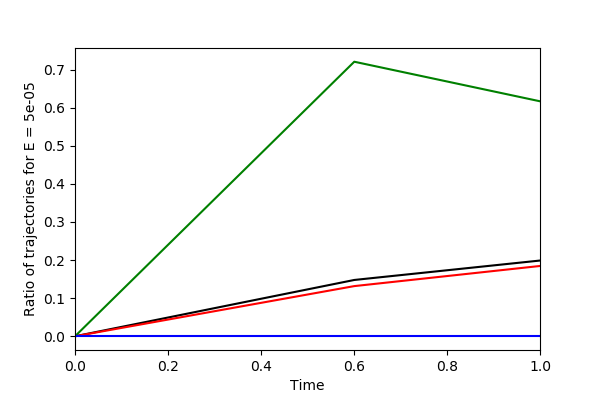}  
        B)\includegraphics[scale = 0.5]{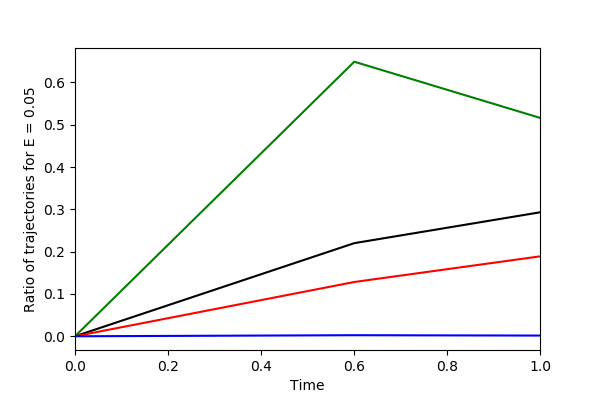}\\
        C)\includegraphics[scale = 0.5]{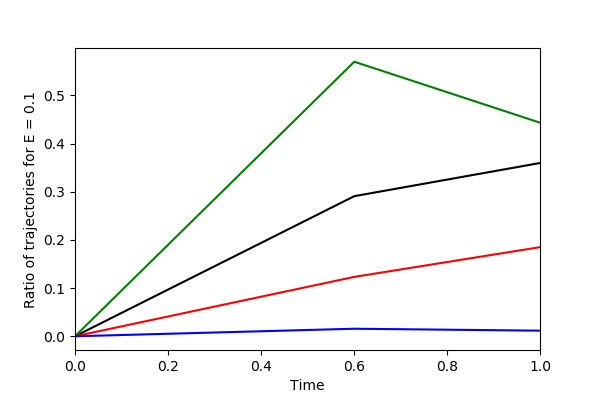}
        D)\includegraphics[scale = 0.5]{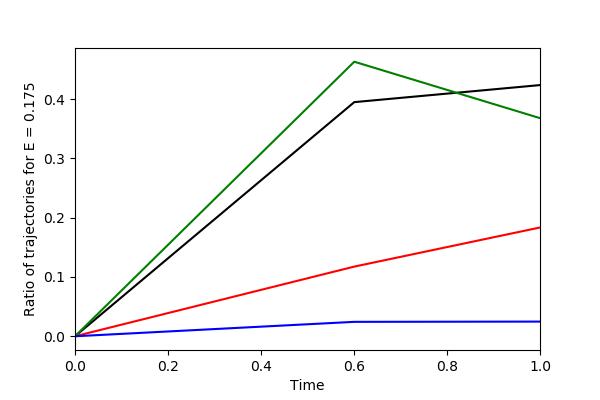}\\
     \end{center}
     \caption{ This is an enlargement of Fig. \ref{tratio1} for the time interval $[0,1]$.}
     \label{tratio1a}
\end{figure}

We integrate 10000 initial conditions on the dividing surfaces of the top unstable periodic orbits (for values of energy below 0.2 (Case I) or for values of energy above 0.2 (Case II)). Half of the initial conditions  correspond to  trajectories that enter the region of the top well (with initial conditions with positive $p_y$). The other half correspond to  trajectories that exit from  the region of the top well (with initial conditions with negative $p_y$).  We consider two cases in our simulations that are dependent on the energy of the system:

\begin{enumerate}
\item {\bf Case I - Low energy:}
In this paragraph we study the case where the energy is below 0.2 and for a small time interval $[0,0.6]$. We see (in all panels of Fig. \ref{tratio1},\ref{tratio1a}) that the  ratio of the trajectories that are  in the region of the top well (black line) is lower than the ratio of the trajectories that are in the exit/entrance channel (green line). The difference between the two ratios is larger for lower values of energy. This means that many of the trajectories that enter into the region of the top well move very fast to the region of the entrance/exit channel. The mechanism for this  transport is, as we described in \cite{katsanikas2020c}, that  the trajectories  follow the unstable invariant manifolds of the top unstable periodic orbits and through heteroclinic intersections they follow the stable invariant manifolds of the unstable periodic orbits of the upper index-1 saddle. Furthermore (for the same time interval), we see  a  small ratio of trajectories to be in the region of the bottom well (red line, see all panels of Fig. \ref{tratio1}). This is because some trajectories that exit from the region of the top well then go to the region of the entrance/exit entrance, and subsequently move very fast to the region of the bottom well. This happens because the trajectories initially follow the unstable manifolds of the  top unstable periodic orbits and through  heteroclinic  intersections they follow the stable invariant manifolds of the  bottom unstable periodic orbits
(see \cite{katsanikas2020c} for more details for this mechanism of transport). The ratio of the trajectories  that are in the  bottom well (red line) is lower than this of the trajectories that are in the top well (black line, see all panels of Fig. \ref{tratio1}).

For larger times, the ratio of the trajectories that are in the top well (black line) appears to have a decreasing trend for a fixed value of energy, with small slope and small fluctuations (see Fig. \ref{tratio1}). This means that only a few trajectories leave this  region. On the contrary, the ratio of the trajectories that are in the bottom well (red line) shows an increasing trend with small fluctuations (see Fig. \ref{tratio1}). Eventually it approximates  the ratio of the trajectories  that are in the top well (black line, see Fig. \ref{tratio1}). This means that  many  trajectories  move from the region of the entrance/exit channel to the region of the bottom well. In addition, we have an increase of the  trajectories  that exit (blue line, see the increasing trend of the ratio of the trajectories that exit - Fig. \ref{tratio1}). This explains, together with the increase of the ratio of the trajectories that are in the bottom well (red line) and the approximately constant rate of the trajectories that are in the top well, the decrease of the ratio of the trajectories that are in the entrance/exit channel (green line, see Fig. \ref{tratio1}). However, we notice that the ratio of the trajectories that are in the entrance/exit channel (green line)  is larger  than the  ratio of the trajectories that are in the other regions, in the final and in most of the medium time intervals (see Fig. \ref{tratio1}).

\item {\bf Case II - High Energy:}
In this paragraph we study the case where the energy is above 0.2 and for a small time interval $[0,0.6]$. We notice that for this small time interval (in all panels of Fig. \ref{tratio2}) the situation is very different from that of case I for the same time interval. We observe that in all panels of Fig. \ref{tratio2} the  ratio of the trajectories that are  in the region of the top well (black line) is larger than that of the trajectories that are in the exit/entrance channel (green line). The difference between the two ratios becomes larger as the energy increases. The reason for this is that many trajectories from those that initially leave the region of the top well move very fast to the region of the bottom well or to the exit (see the increase of the trajectories that are in the bottom well (red line) or that exit (blue line) in all panels of Fig. \ref{tratio2}). This means that many trajectories that initially left the region of the top well follow the unstable invariant manifolds of the top unstable periodic orbits until they follow, through heteroclinic intersections (see \cite{katsanikas2020c}), the stable invariant manifolds of the bottom unstable periodic orbits to the region of the bottom well.  Furthermore, many of the trajectories that initially left the region of the top well follow the unstable invariant manifolds of the top unstable periodic orbits until they follow, through heteroclinic intersections (see \cite{katsanikas2020c}), the stable invariant manifolds of the  unstable periodic orbits of the upper index-1 saddles to the exit. 

For larger times the ratio of the trajectories that are in the top well (black line) has a decreasing trend for a fixed value of energy, initially with a large slope and then with a small slope and small fluctuations (see Fig. \ref{tratio2}). In addition, the ratio of the trajectories that are in the  entrance/exit channel (green line), that initially increases, is approximately constant (if we compare the ratio of these trajectories for Time = 1 with Time = 6 in Fig. \ref{tratio2}) despite the small fluctuations (see Fig. \ref{tratio2}). On the contrary, the ratio of the trajectories,  that are in the bottom well (red line) or that exit, shows an increasing trend with small fluctuations (see Fig. \ref{tratio2}). This means that we have  transport of trajectories from the region of the top well  to the region of the bottom well and from the region of the top well to the exit.  This  happens because many trajectories that are in the region of the top well  follow initially the unstable invariant manifolds of the top unstable periodic orbits until they become guided by (through heteroclinic intersections)  the stable invariant manifolds of the bottom unstable periodic orbits or of the unstable periodic orbits of the upper index-1 saddle (see \cite{katsanikas2020c}). This transport is very fast  since it does not affect the ratio of the trajectories that are in the entrance/exit channel but directly the ratios of the trajectories that are in the bottom well or that exit. We note that we did not observe  such fast transport like this in  case I. 

\end{enumerate}

\newpage

\newpage

\section{Conclusions}
\label{concl.}

We studied the time evolution of the trajectories  after selectivity in a symmetric system with two sequential index-1 saddles and two wells. We  investigated the time evolution  of the trajectories that have initial conditions on the periodic orbit dividing surfaces of the top unstable periodic orbits, as energy varies (the results can be obtained in a similar way for the dividing surfaces of the  bottom unstable periodic orbits because of the symmetry of the potential). In this way, we studied the  trajectories that enter into,  or exit from, the regions of the wells. This is very important because we investigate the destiny of the trajectories after the selectivity. Our conclusions about the destiny of the trajectories is summarized in the next two cases that are dependent on the value of energy: 

\begin{enumerate}

    \item {\bf Case I -Low energy} (for values of energy below 0.2): For lower values of energy the ratio of the trajectories in the entrance/exit channel is initially  larger than this of the trajectories of the wells. For time larger than 0.6, many trajectories in the entrance/exit channel move to the region of the wells or to the exit. Despite this fact, the ratio of the trajectories in the entrance/exit channel  remains always larger than that of the trajectories of the wells.
    
    \item {\bf Case II - High energy} (for values of energy above 0.2):  For higher values of energy the ratio of the trajectories of the entrance/exit channel is initially  smaller  than this of the trajectories of the wells. For time larger than 0.6, many trajectories leave the region of each of the two wells and  visit the region of the other well,  or they exit. 
    In this case, we have  fast transport from the region of one of the two wells to the other well, or to the exit. In addition,  the ratio of the trajectories at the entrance/exit channel intially increases but  then it  remains approximately constant, with some fluctuations.

 \end{enumerate}

\section*{Acknowledgments}

The authors would like to acknowledge the financial support provided by the EPSRC Grant No. EP/P021123/1.


\end{document}